\documentclass[conference]{IEEEtran}
\usepackage{graphicx}
\usepackage{caption}
\usepackage{subcaption}
\usepackage{color}
\usepackage{amssymb}
\ifCLASSINFOpdf
\else
\fi
\usepackage{mathtools}
\DeclarePairedDelimiter\ceil{\lceil}{\rceil}

\usepackage{amsmath}
\usepackage{algorithmic,algorithm}
\allowdisplaybreaks
\begin{document}

\title{Learning-Based Link Scheduling in Millimeter-wave Multi-connectivity Scenarios}

\author{\IEEEauthorblockN{Cristian Tatino\IEEEauthorrefmark{1}, Nikolaos Pappas\IEEEauthorrefmark{1}, Ilaria Malanchini\IEEEauthorrefmark{2}, Lutz Ewe\IEEEauthorrefmark{2}, Di Yuan\IEEEauthorrefmark{1}}
\IEEEauthorblockA{\IEEEauthorrefmark{1}Department of Science and Technology, Link\"{o}ping University, Sweden\\
               Email: \{cristian.tatino, nikolaos.pappas, di.yuan\}@liu.se}      
               \IEEEauthorblockA{\IEEEauthorrefmark{2}Nokia Bell Labs, Stuttgart, Germany\\
               Email: \{ilaria.malanchini, lutz.ewe\}@nokia-bell-labs.com}
}

\maketitle
\IEEEpeerreviewmaketitle
\begin{abstract}
Multi-connectivity is emerging as a promising solution to provide reliable communications and seamless connectivity for the millimeter-wave frequency range. Due to the blockage sensitivity at such high frequencies, connectivity with multiple cells can drastically increase the network performance in terms of throughput and reliability. However, an inefficient link scheduling, i.e., over and under-provisioning of connections, can lead either to high interference and energy consumption or to unsatisfied user's quality of service (QoS) requirements. In this work, we present a learning-based solution that is able to learn and then to predict the optimal link scheduling to satisfy users' QoS requirements while avoiding communication interruptions. Moreover, we compare the proposed approach with two base line methods and the genie-aided link scheduling that assumes perfect channel knowledge. We show that the learning-based solution approaches the optimum and outperforms the base line methods.
\end{abstract}

\section{Introduction}
\label{sec:Intro}
Ever increasing traffic demand leads telecom operators to constantly improve spectrum efficiency in the available bandwidth and to find new spectrum resources. Thus, the millimeter-wave (mm-wave) frequency range (30-300 GHz) provides a big opportunity to enable high throughput transmissions and increase the number of served users. However, such high frequencies are characterized by more challenging propagation conditions than lower frequency bands, especially in terms of free space path loss and penetration loss~\cite{BLO}. The use of high directional beams can partially solve these issues by providing high beamforming gains. On the other hand, narrow beams introduce further complexity and delay when communications need to be established or recovered~\cite{OurRel}. Moreover, due to the use of narrow beams and the high path loss, the signal arrives at the receiver along few paths~\cite{Meas}, which can be easily blocked due to the high penetration loss causing frequent communication interruptions.

Several solutions have been proposed in order to minimize communication interruptions, e.g., multi-connectivity (MC). This allows a user to establish and maintain connections with multiple cells/access points at the same time. Mm-wave wireless networks can be also assisted by low-band frequencies that support control-plane traffic and, when mm-wave links are not available, data transmissions~\cite{Danish}. In this work, we propose a learning-based solution for link scheduling optimization in low-band assisted mm-waves MC scenarios. This is a critical problem for which the solution is not obvious, since it should optimize the tradeoff between increasing throughput and reliability when using more links and decreasing the energy consumption and interference caused. Thus, network performance are highly affected by the number and the selection of links that depend on several parameters, e.g., type of user (vehicle or pedestrian), quality of service (QoS) requirements, and blockage frequency. Our proposed solution is able to learn and then to predict the optimal choice of links that are necessary in order to serve the user with the required QoS, while minimizing the cost of allocated resources, and avoiding those links that may be subject to blockages.

\subsection{Related Works}
\label{sec:Rel}
Learning techniques applied to wireless networks are gaining increasing attention. In mm-wave frequency band, learning methods have been applied mainly to mobility management and initial access, e.g., \cite{RLH,NNH,MLB,DLCB}. In~\cite{RLH}, the authors propose a reinforcement learning solution for minimizing the number of handovers. A similar problem is analyzed in~\cite{NNH}, where the authors use a neural network that takes into account the past sequence of beams for predicting handovers in urban environments in order to avoid blockages. However, these works consider a single connectivity and static scenario where moving obstacles are disregarded. 

Moving obstacles are considered in~\cite{MLB,DLCB}. In the former, several supervised learning models are used to investigate beam-selection techniques on mm-wave vehicle-to-infrastructure networks. Mm-wave multi-connectivity scenarios are considered in~\cite{DLCB}, where the authors propose a supervised learning method for coordinated beamforming. This solution consists of a deep neural network that is able to predict the best beam for multiple millimeter-wave access points by using only omnidirectional uplink pilot signals. However, the proposed solution is used for initial access purposes and does not minimize allocated resources since it does not consider any QoS requirement. Link scheduling optimization for mm-wave multi-connectivity scenarios is considered in~\cite{Our1}, however the proposed solution assumes perfect channel knowledge and a similar approach can be used to compute an upper bound for the performance of the learning-solution proposed in this work.

\subsection{Contributions}
\label{sec:Contributions}
In this work, we propose a novel learning-based solution for link scheduling in multi-connectivity scenarios for mm-wave wireless networks. The proposed method is a \textit{random forest classifier}~\cite{ISL} that is able to learn the optimal selection of links with respect to users' location and QoS requirements in order to achieve the following goals: i) satisfying users' QoS requirements, ii) minimizing the allocated resources, and iii) offloading low frequency communications. Namely, this last objective is pursued due to the scarcity of spectrum resources at low frequency bands, which are intended to provide radio coverage and serve users that cannot be served by any mm-wave access point. We show the performance of the learning-based model for several mobility types, i.e., pedestrians and vehicles. Moreover, we compare the obtained results with a genie-aided solution and two base line methods. The performance evaluation shows that the random forest classifier is able to outperform all the considered base line methods and, for some scenarios, it approaches the optimum.

The rest of the paper is organized as follows: In Section II we describe the system model and the assumptions. In Section~III and in Section~IV we formulate the optimization problem and present the learning solution, respectively. Finally, Section~V illustrates the results and performance comparison and Section VI concludes the paper.

\section{System Model}
\label{sec:Ass}

We consider a single user downlink outdoor scenario of a low-band assisted mm-wave wireless network, that consists of a set of cells $\mathcal{C}$, with cardinality $|\mathcal{C}|=N+1$ and namely composed by $N$ mm-wave access points (mmAPs), and one low-band base station (LB-BS). Hereafter, we use index $i=0$ and $i=1,...,N$ to indicate the LB-BS and the mmAPs, respectively. As in~\cite{Danish}, the mmAPs and LB-BS jointly provide coverage and connectivity to a mobile user in a multi-connectivity scheme and they are connected by high speed backhauling links to a network controller (NC). Between each cell $i \in \mathcal{C}$ and the user there is a wireless link that can be used for downlink packet transmissions. We assume that mmAPs transmit on different frequencies and the user has multi-packet reception capability. Moreover, we assume slotted time and each packet transmission takes one timeslot.

\begin{figure}[tb]
	\centering
	\includegraphics[width=8cm]{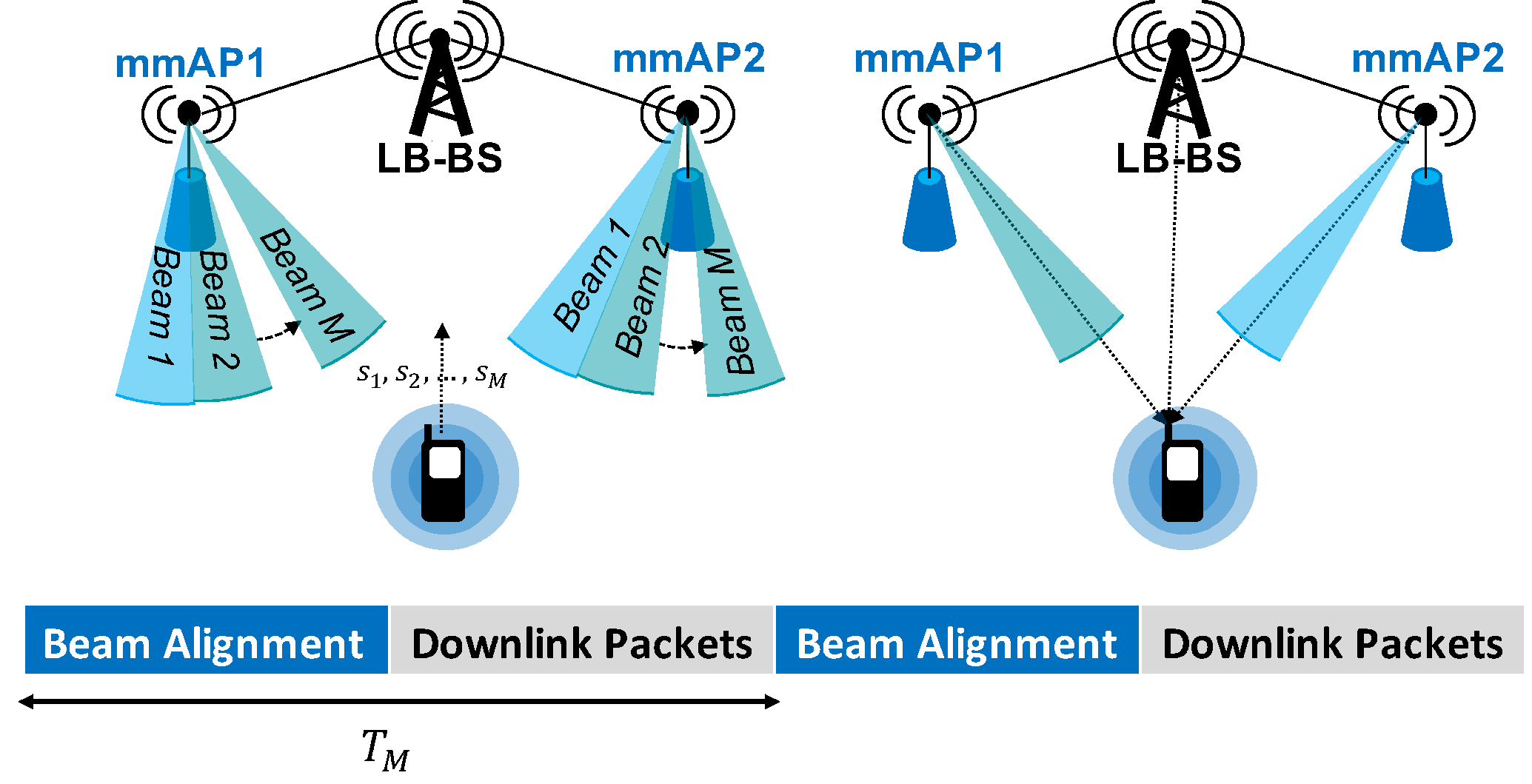}
	\caption[]{Low-band assisted mm-wave wireless network with two mm-wave access points (mmAPs), and a low-band base station (LB-BS). In the left, an uplink beam alignment strategy it is shown, where the user transmits $M$ pilot signals $s_{j}, j=1,...,M$. In the right, two mmAPs transmit packets to the user by using the beam with the highest SNR in a multi-connectivity scheme.}
	 \label{fig:scen}
\end{figure}

As shown in Fig.~\ref{fig:scen}, every $T_{M}$ timeslots a beam alignment phase occurs, which is followed by downlink packet transmissions. During the alignment phase the user  transmits a certain number of pilots, which are received by mmAPs on different beams. This is an uplink strategy, as proposed in~\cite{UPM}, that decreases the complexity of the beam alignment phase for MC scenarios. Moreover, we assume that mmAPs are equipped with uniform planar antenna arrays that can form a set of $M$ 3-dimensional beams, whereas the user is equipped with an omnidirectional antenna. Once NC obtains signal-to-noise (SNR) measurements for each beam and each mmAP, it schedules packet transmissions for the next scheduling window on one or multiple links (mmAPs and LB-BS). These are selected in order to satisfy the user's QoS requirements. We assume that the mmAPs transmit by using the beam that has the highest SNR. QoS requirements are stated in terms of amount of data that the user needs to receive within a deadline, i.e., $D$ packets in $K$ timeslots, that the user needs to receive. The NC can exploit multiple links either to transmit different packets simultaneously, or, by exploiting spatial diversity and applying packet duplication, transmit the same packet over multiple links~\cite{MCRT}. In this work, we consider the first strategy and a packet is successfully received if the SNR at the receiver is higher than a threshold $\gamma$, i.e., SNR$\ge \gamma$. If a transmission fails, the packet is retransmitted.

\subsection{Channel Model}
\label{sec:channel}
In order to obtain the SNR for each beam and mmAP at the user position, we adopt a wideband multi-path mm-wave channel model~\cite{CHM} with $L$ clusters. Each cluster $l=1,...,L$ is assumed to contribute with one ray that has a time delay $\tau_{l}$, azimuth and elevation angles of arrival $\theta_{l,r}$ and $\phi_{l,r}$, and azimuth and elevation angles of departure $\theta_{l,t}$ and $\phi_{l,t}$ respectively. Moreover, we denote the complex gain and the Doppler shift of the $l$-th ray between the user and the mmAP with $\alpha_{l}$ and $\nu_{l}$, respectively. Then, we can write the received signal as follows:
\begin{align*} 
y(t)&=\sum_{l=1}^{L}\alpha_{l}e^{j2\pi \nu_{l}t}a_{t}(\theta_{l,t},\phi_{l,t})a_{r}(\theta_{l,r},\phi_{l,r})x(t-\tau_{l})+v(t) \stepcounter{equation}\tag{\theequation}\label{eq:cha},
\end{align*}
where, $x(t)$, $y(t)$ and $v(t)$ are the transmitted, received, and noise signals, respectively. Moreover, $a_{t}$ and $a_{r}$ are the array response vectors at the mmAP and at the user, respectively. In order to obtain the parameters that completely characterize a single ray, i.e., $\alpha_{l}$, $\tau_{l}$, $\nu_{l}$, $\theta_{l,t}$, $\phi_{l,t}$, $\theta_{l,r}$, and $\phi_{l,r}$, we use a ray tracing simulator. This enables the computation of the received signal power and the received SNR for each beam and user location. Moreover, we consider a blockage model for which a ray is blocked if there is a moving or a fixed obstacle along the ray path between the transmitter and the receiver. More details about the ray tracing simulator, obstacle characteristics, and the adopted antenna model are provided in Section~\ref{sec:PerfEv}.

\section{Problem Definition and Base Line Methods}
\label{sec:Prob}
The network controller allocates the links for downlink transmissions while fulfilling the following objectives:
\begin{itemize}
\item Satisfy the user's QoS requirements in terms of amount of successfully received packets within the corresponding deadline
\item Minimize the number of failed transmissions
\item Offload LB-BS
\end{itemize} 
In this scheduling problem, a challenge to deal with is that the channel conditions for $k=1,...,K$ are not known, as only the measurements resulting from the alignment phase at $k=0$ are available. However, using solely these is not enough to perform optimal link scheduling, since mm-wave links are frequently blocked by fixed or moving obstacles that cause failed transmissions. Contrary to fixed obstacles, moving objects, e.g., vehicles and pedestrians, are difficult to predict. When the risk of blockages is high, packets should be transmitted by using multiple paths. However, an inefficient link scheduling may select links that are going to be subject of blockages leading to frequent failed transmissions. These can lead to higher energy consumption and unsatisfied users. 

The NC can learn that some locations are particularly crowded, i.e., a street, a bus stop, a metro station exit, and link blockages are more frequent, thus more links need to be allocated to the user in order to guarantee the required QoS. For this reason, we propose a learning-based solution that is able to learn the optimal link scheduling for $K$ future slots according to QoS requirements and user position. Moreover, in order to evaluate the performance of our solution, in the following sections, we propose a genie-aided link scheduling optimization, which assumes perfect future channel knowledge, as well as two baseline methods.

\subsection{Genie-aided Link Scheduling}
\label{sec:Opt}
In this section, we formulate an integer optimization problem (IP) that assumes perfect knowledge of the channel for $k=1,...,K$. First, we define a set $\mathcal{S}$ of link scheduling strategies with cardinality $|\mathcal{S}|=2^{N+1}$ that contains all the possible combinations of $N+1$ links ($N$ mmAPs and LB-BS). Each combination is represented by a set $\mathcal{V}_{j}$, with $j=1,...,|\mathcal{S}|$, which contains the links that are scheduled for the user. For each combination of links $\mathcal{V}_{j}$, we define a variable $z^{j}$ that is equal to $1$ if the $j$-th link scheduling strategy is used and $0$ otherwise. Then, for each timeslot $k=1,...,K$ and for each cell $i \in \mathcal{C}$, we define a binary variable $y^{ik}$, which is equal to $1$ if at timeslot $k$ a packet is transmitted by using cell $i$ and $0$ otherwise. Clearly, if NC selects the $j$-th scheduling strategy in $\mathcal{S}$, packets can be transmitted only by using links in $\mathcal{V}_{j}$. Finally, given the amount of packets to be transmitted, $D$, the IP can be formulated as follows:
\begin{subequations}
\begin{align}
        IP:&\min_{y^{ik},z^{j}} \sum_{k=1}^{K}\sum_{i=0}^{N}c_{i} y^{ik}\label{opt}\\ 
        \text{s.t.}&\sum_{k=1}^{K}\sum_{i=0}^{N} g_{ik}y^{ik}\ge D, \label{D_con}\\
        		&\sum_{j=1}^{|\mathcal{S}|}z^{j} = 1, \label{Comb}\\
		&\sum_{j:i\in \mathcal{V}_{j}}z^{j}  \ge y^{ik}, \quad \forall i\in\mathcal{C}, k=1,...,K \label{Act}\\
		 &y^{ik},z^{j} \in\{0,1\} \label{type} \quad \forall i\in\mathcal{C}, k=1,...,K, j=1,...,|\mathcal{S}|.
\end{align}
\end{subequations}
The objective function, given by~\eqref{opt}, represents the sum among all the links of the allocated timeslots for the packet transmissions, where $c_{i}$ is the cost to transmit a packet on link $i$. In our setup, we set $c_{i}=1, \forall i \in\mathcal{C}\setminus\{0\}$, and, in order to offload the low frequency cell, we set $c_{0} \gg c_{i}\ge 1, \forall i \in\mathcal{C}\setminus\{0\}$. Constraint~\eqref{D_con} takes into account user QoS requirements and imposes the number of successfully received packets over $K$ timeslots to be at least $D$. Binary coefficients $g_{ik}, \forall i\in\mathcal{C}, k=1,...,K$, represents the channel condition for the link between the $i$-th cell and the user at timeslot $k$. Namely, $g_{ik}$ is equal to 1 when the received SNR for such link is higher than the threshold $\gamma$. Recall that, the packet transmission between the $i$-th mmAP and the user  occurs through the beam that results from the alignment. This is the beam with the highest received SNR that is used to set the coefficient $g_{ik}$. Constraint \eqref{Comb} imposes that exactly one combination of links is selected. Finally, constraints \eqref{Act} identify on which links packet can be transmitted according to the selected combination and state the consistency between the $z$-variables and $y$-variables. Note that for the scope of the genie-aided formulation, having both $z$ and $y$ variables is not strictly necessary. However we introduced $z$ here, since this will be used for training the proposed learning-based approach.


\subsection{Base Line Methods}
\label{sec:heur}
In this section, we provide two heuristics that do not assume perfect channel knowledge: i) Greedy Multi-x, and ii) Min Multi-x. The former represents a greedy solution that consists of transmitting packets, in each time slot, by using all the available mmAPs and LB-5G links. This provides an upper bound of the cost and completed tasks. On the other hand, Min Multi-x allocates $\ceil*{D/K}$ links by selecting those (mmAPs or LB-BS) with the highest SNR measurements, where $\ceil*{x}$ represents the ceiling function that maps $x$ to the smallest integer greater than or equal to $x$. Thus, Min Multi-x transmits packets by using the minimum number of links that are necessary to satisfy the QoS requirements.

\section{Learning-based Link Scheduling}
\label{sec:Learn}
\begin{figure}[tb]
	\centering
	\includegraphics[width=7cm]{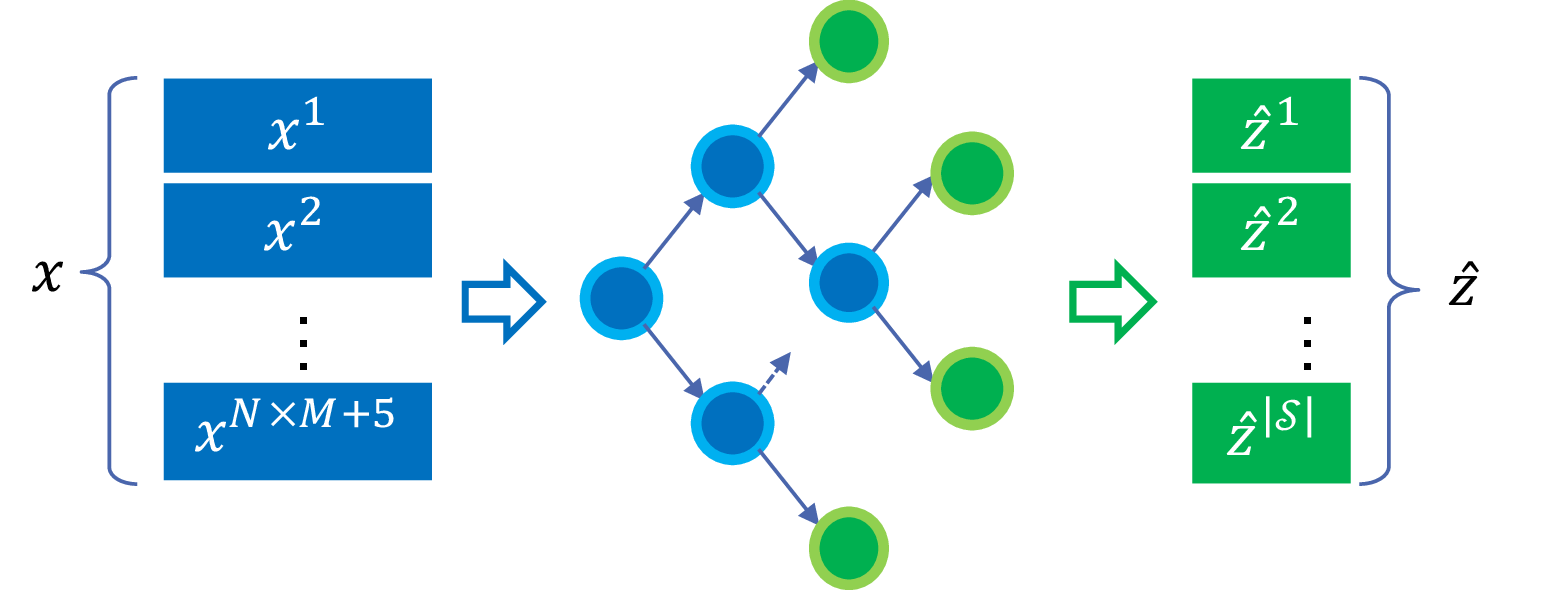}
	\caption[]{Random forest for link scheduling; $x$ and $\hat{z}$ are the input and output vectors, respectively. We show the internal and terminal nodes in blue and green, respectively.}
	 \label{fig:tree}
\end{figure}

To overcome the unrealistic assumption of perfect channel knowledge, we propose a learning-based solution that is able to learn the optimal link scheduling strategy for $K$ future slots for different QoS requirements. Namely, as it is shown in Fig.~\ref{fig:tree}, we consider a random forest classifier that, given an input vector $x \in R^{N \times M + 5}$, provides an output binary vector $\hat{z} \in \{0,1\}^{|\mathcal{S}|}$ corresponding to the chosen link strategy.
The input vector $x$, consists of the following elements:
\begin{itemize}
\item $N \times M + 1$ elements corresponding to the SNR measurements for the LB-BS and for each beam of the mmAPs.
\item 2 elements representing the number of packets to be transmitted $D$ and the corresponding deadline $K$.
\item 2 elements corresponding to the user position coordinates. 
\end{itemize} 
Namely, the $j$-th element of the output vector $\hat{z}$, represents an estimation of the variable $z^{j}$ of IP, and it is equal to $1$ if the $j$-th combination of links is used to schedule the packet transmissions for the user and $0$ otherwise. In addition to $\hat{z}$, the random forest provides a vector $\hat{p}$, whose $j$-th element, $0 \le \hat{p}^j \le 1$, represents the probability that combination $j \in \mathcal{S}$ is the optimal scheduling strategy. This probability can be used to measure the uncertainty of the predicted solution $\hat{z}$. 
\subsection{Training and Prediction Phases}
\label{sec:channel}
We use a supervised learning approach to train our classifier. We consider a dataset composed by $E$ training samples, each consisting of an input vector $x_{e}$, and a corresponding output vector $z_{e}$, with $e=1,...,E$. The latter represents the optimal link scheduling solution that is obtained by solving a modified version of IP in Section~\ref{sec:Opt}. More precisely, we add constraints to IP to force to transmit consecutively in each timeslot and on each link that is identified by the optimal strategy $z_{e}$ until either $k=K$ or all the packets have been transmitted. We introduce such constraints because this is also how the NC operates during the prediction phase, as is explained below. To solve the modified version of IP, NC needs to know $g_{ik}$ for $i\in\mathcal{C}$ and $k=1,...,K_{e}$. Recall that, $g_{ik}=1$ if the received SNR for the $i$-th link at timeslot $k$ is higher than the threshold $\gamma$. Thus, these coefficients can be obtained by transmitting on all the links in each timeslot, then $g_{ik}=1$ if the transmitted packet on link $i$ at timeslot $k$ is successfully received, and $g_{ik}=0$ otherwise. Then, NC is able to compute the solution, $z_{e}$, by solving the modified IP. In case this is infeasible, $z_{e}$ corresponds to the decision not to transmit packets. Once the dataset is obtained, it is used to train our random forest. This is composed by several decision trees, whose results are averaged. During the training phase, each decision tree splits the input space into several regions that correspond to the leaf nodes. The training process has the objective to minimize the Gini index~\cite{ISL}.

We now describe how our learning solution is used to schedule the downlink packet transmissions. This is called prediction or scheduling phase. More specifically, once the NC obtains measurements and position of the user (every $T_{m}$ timeslots), these form, with user's QoS requirements ($D$ and $K$), vector $x$. This is given as input to the random forest that provides output vectors $\hat{z}$ and $\hat{p}$. Then, in each timeslot, downlink packets are transmitted by using the links belonging to the $j$-th combination in $\mathcal{S}$ if and only if $\hat{z}^{j}=1$ and $\hat{p}^j$ is higher than a threshold $0 \le \beta \le 1$, i.e., $\hat{p}^j> \beta$. Recall that $\hat{p}^j$ represents the probability that combination $j \in \mathcal{S}$ is the optimal scheduling strategy. On the other hand, if $\hat{p}^j \le \beta$, NC decides to transmit packets by using all the available links. This strategy represents the solution that is taken by Greedy Multi-x, described in Section~\ref{sec:heur}, and it is characterized by a higher probability to satisfy the user's QoS requirements, but with higher costs. Namely, $\beta$ represents a threshold on the uncertainty of the predicted solution of the random forest and it is selected in order to guarantee a certain level of performance. 

 \begin{figure}[tb]
	\centering
	\includegraphics[width=7cm]{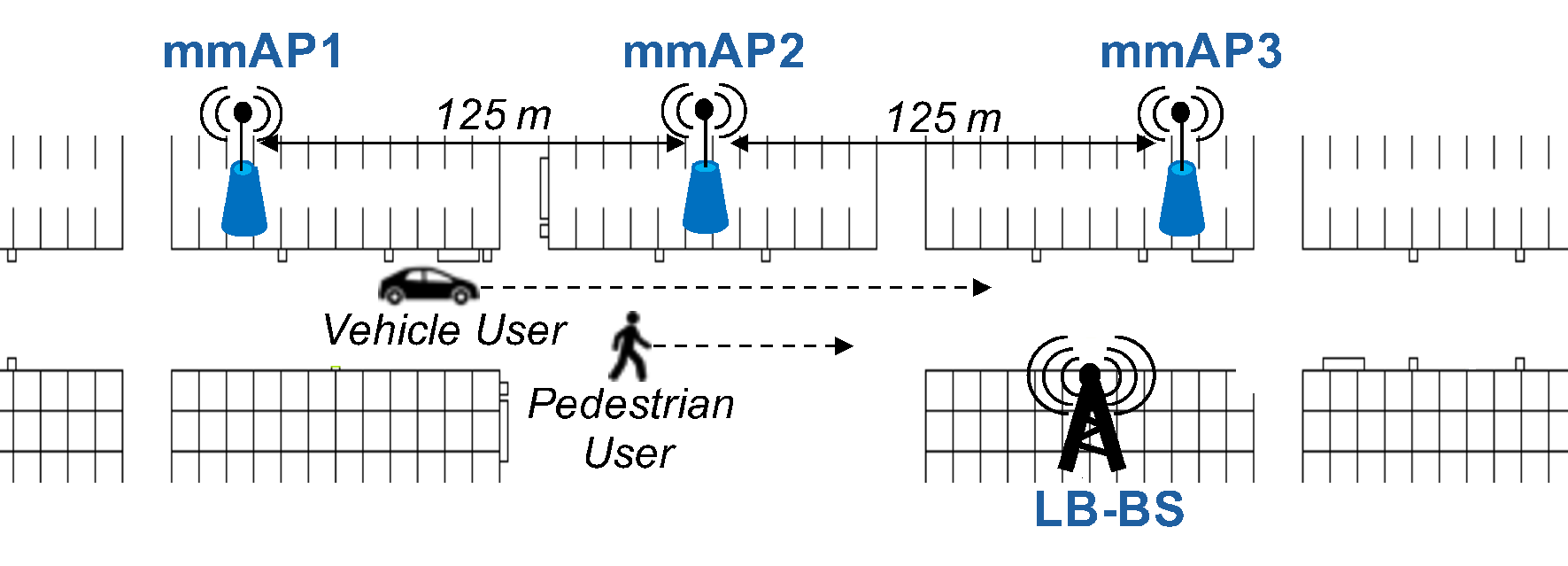}
	\caption[]{Dense urban area from ``Madrid Grid" scenario, where, a vehicle and a pedestrian user move along a straight path inside to the coverage area of $3$ mmAPs and one LB-BS.}
	 \label{fig:Sim}
\end{figure}

\section{Performance Evaluation}
\label{sec:PerfEv}
In this section, we provide a numerical evaluation of the random forest\footnote{We evaluated also alternative learning solutions, such as neural networks and k-nearest neighbors, but numerical results showed that random forest outperforms all of them.}, the genie-aided, and the two base-line solutions described in the previous sections. For our simulations we consider $N=3$ mmAPs, each operating at a different frequency channel in the $28$~GHz band, and one LB-BS operating at $2$ ~GHz. These are located as shown in Fig.~\ref{fig:Sim}, in a subarea of the ``Madrid Grid" scenario, as defined by project METIS~\cite{MET}, which is an example of a dense urban area with significant shadowing by high buildings separated by street canyons. Both mmAPs and LB-BS are placed at a height of $10$\,m, whereas, we set the antenna height at the users equal to $1.5$\,m. The transmit and the noise powers are set to $24$\,dBm and $-80$\,dBm. The antenna at the mmAPs is configured as an $8\times8$ array of antenna elements generating a pattern of $19$ horizontal fixed beams with a pattern gain of $18$ dB~\cite{ANT} at a single elevation (i.e., down tilt) angle of $8$ degrees. It covers a total angular range of $180$ degrees, i.e. leading to a $10$-degree single beam step. The SNR threshold is set to $\gamma = 10$ dB.

In order to obtain the received power and the SNR for each user location we use a ray tracing simulator, which takes into account fixed (e.g., buildings) and moving obstacles. We consider three types of moving obstacles, i.e., large (e.g., buses) and small vehicles ( e.g., cars), and pedestrians, whose starting positions are randomly dropped in the considered area with densities of $0.03$/m$^2$ for pedestrians and $0.01$/m$^2$ for vehicles. Then, starting from these positions, they move along a straight path at a constant speed of $3$\,Km/h, $30$\,Km/h, and $50$\,Km/h for pedestrians, large and small vehicles, respectively. Vehicles have rectangular shapes with width of $2.2$\,m, and with length of $8$\,m and $4$\,m for large and small ones, respectively. Their heights are set to $3$\,m and $1.8$\,m, whereas, for pedestrians, we assume an average height of $1.75$\,m.

We train and the test our model on two types of users, i.e., a small vehicle and a pedestrian. These move for $10$\,s along a straight path, as shown in Fig.~\ref{fig:Sim}. Then, we consider $500$ realizations, each consisting of $10,000$ locations with related SNR measurements, with a timeslot duration of $1$\,ms. Then, for a scheduling window and QoS deadline of $K$, the training set consist of $5,000,000/K$ samples, i.e., for a deadline of $K=50$ timeslots, we obtain $100,000$ training samples for each type of user. For training and testing the random forest we use \textit{scikit-learn} libraries~\cite{scikit} with \textit{TensorFlow} backend. The random forest is composed by $200$ trees. For each of them, we set a maximum depth of $20$ leaf nodes. Finally, we set $K=50$, $D=100$, and test our model on $10$ realizations, as described above, and average the results. Each realization is composed by $200$ scheduling windows (episodes), for a total of $10$\,s. For each episode, the random forest takes approximately $1$\,ms to perform the prediction phase by using a laptop with 8 GB of RAM and a 7th generation, Intel Core i7 processor.
 
\begin{figure}[tb]
	\centering
	\includegraphics[width=6.95cm]{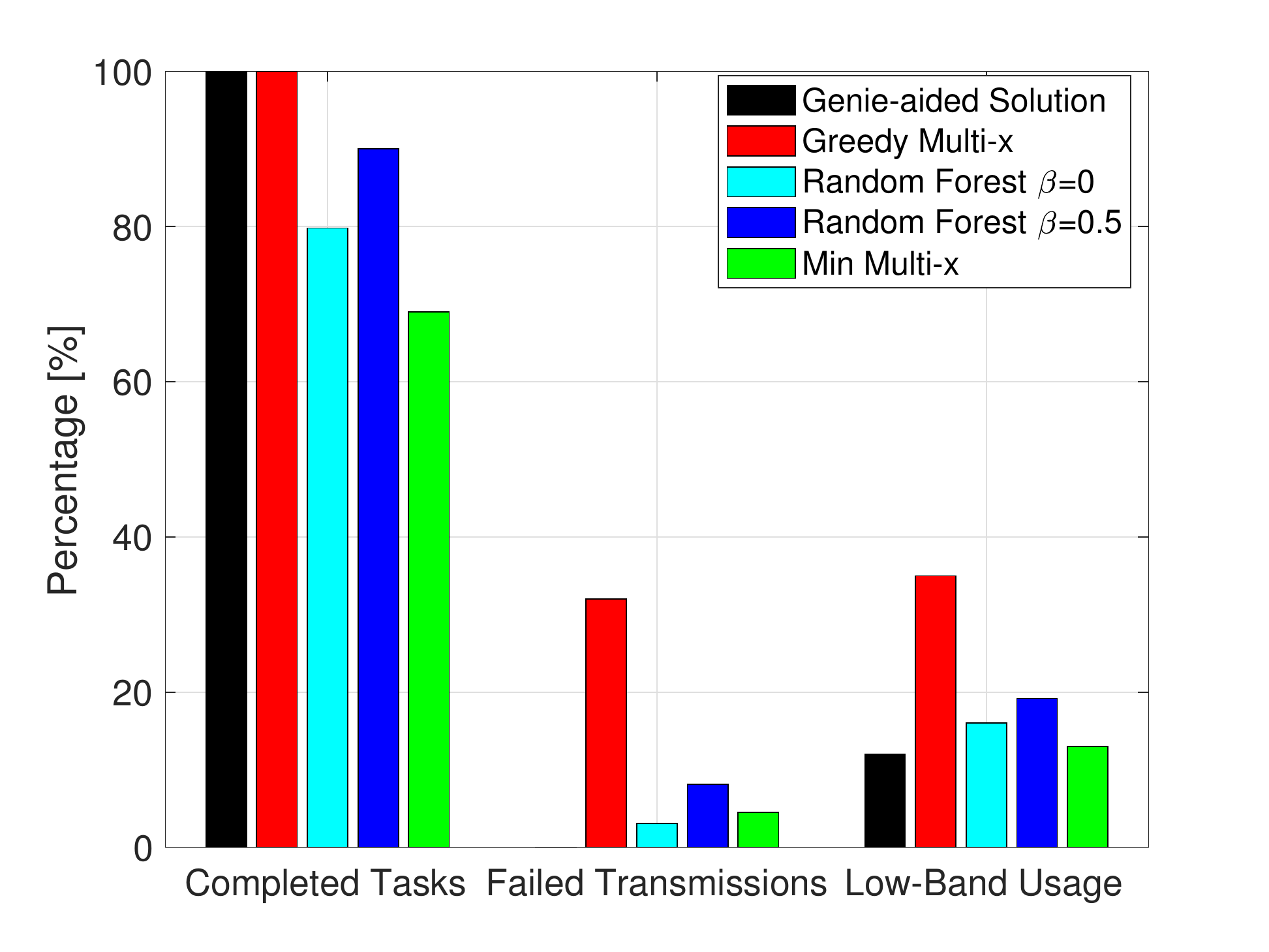}
	\caption[]{Percentage of completed tasks, failed transmissions, and low-band usage for a vehicle user with $D=100$, $K=50$.}
	 \label{fig:100V}
\end{figure}

In Fig.~\ref{fig:100V}, we consider a small vehicle user and show the results for the genie-aided solution, the proposed random forest with $\beta=0$ and $\beta=0.5$, and the two base-line methods (Greedy Multi-x and Min Multi-x). We consider three key performance indicators (KPIs): i) the fraction of episodes for which the QoS requirements are satisfied (completed tasks), ii) the fraction of failed transmissions, and iii) the fraction of packets that are transmitted by using low frequencies (low-band usage). Recall that a packet transmission fails when the received $SNR < \gamma$, and when this happens, the packet is retransmitted. In Fig.~\ref{fig:100V}, we can observe that our solution is able to successfully complete a higher number of episodes than Min Multi-x with a lower fraction of failed transmissions. Moreover, the random forest is able to offload most of the data from LB-BS with a low-band usage that approaches the optimal one. Recall that, Greedy Multi-x uses all the available links. Thus, it is able to complete $100\%$ of the tasks, but with an extremely high cost in terms of failed transmissions and low-band usage. For $\beta=0.5$, the random forest is able to significantly increase the fraction of completed tasks. However, higher values of $\beta$ correspond to higher costs. 

\begin{figure}[tb]
	\centering
	\includegraphics[width=6.95cm]{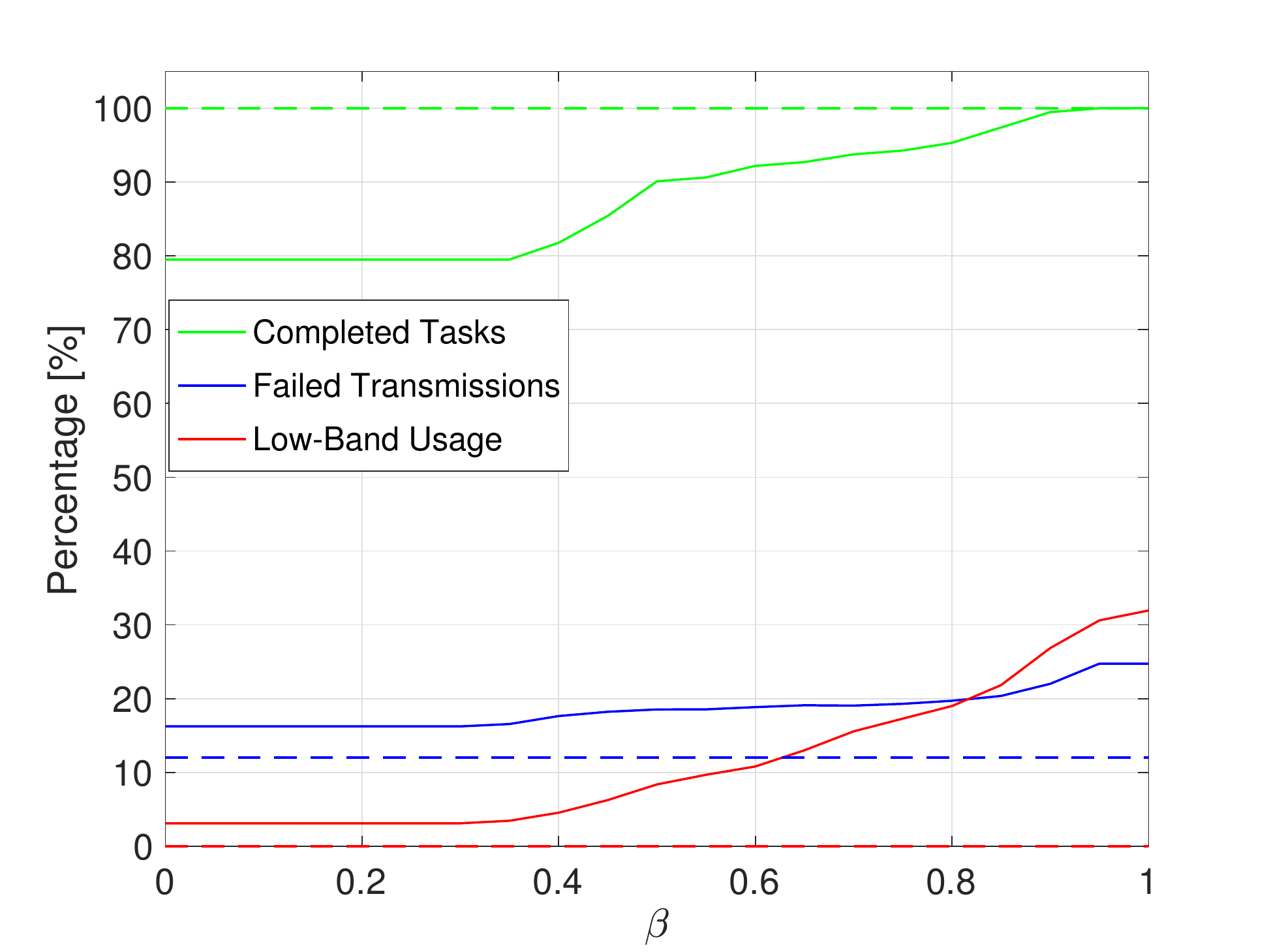}
	\caption[]{Percentage of completed tasks, failed transmissions, and low-band usage with varying $\beta$ for a vehicle user with $D=100$ and $K=50$. We show the results for both the random forest (solid lines) and the genie-aided solution (dashed lines).}
	 \label{fig:Thre100V}
\end{figure}

This is better observed in Fig.~\ref{fig:Thre100V}, where we show with solid lines the KPIs for the random forest while varying the threshold $\beta$ for the same scenario that is analyzed in Fig.~\ref{fig:100V}. With dashed lines we show the performance of the genie-aided solution that does not depend on $\beta$. First, we can observe that for $\beta \le 0.4$, the random forest is able to outperform the Min Multi-x solution without increasing failed transmissions and low-band usage. On the other hand, higher values of $\beta$, corresponds a higher probability to use the fallback solution that leads to transmit packets by using all the links. Despite this solution increases the fraction of completed tasks, it can also increase failed transmissions and low-band usage. The value of $\beta$ is chosen accordingly to the applications and performance required, e.g., in the previous case we set $\beta=0.5$ to have at least the $90\%$ of completed tasks.

\begin{figure}[tb]
	\centering
	\includegraphics[width=6.95cm]{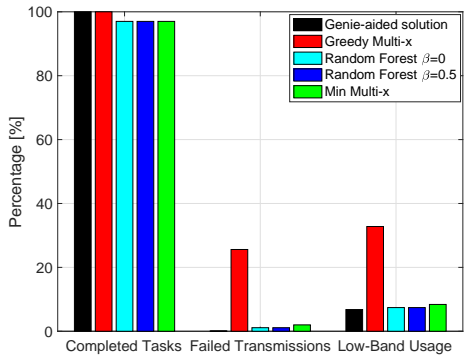}
	\caption[]{Percentage of completed tasks, failed transmissions, and low-band usage for a pedestrian user with $D=100$ and $K=50$.}
	 \label{fig:100Pack_bar}
\end{figure}

\begin{figure}[tb]
	\centering
	\includegraphics[width=6.95cm]{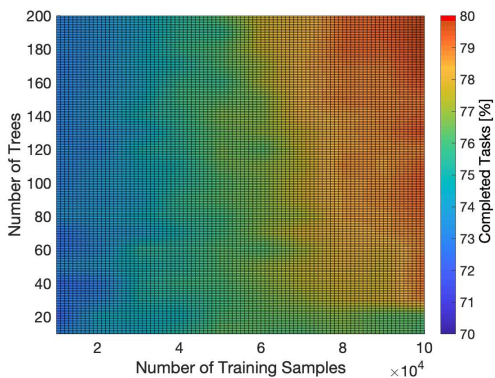}
	\caption[]{Percentage of completed tasks on the test set for a vehicle user with $D=100$, $K=50$, and $\beta=0$, while varying the number of samples on which our model is trained, and the number of trees that compose the random forest.}
	 \label{fig:100_Train_Trees}
\end{figure}

In Fig.~\ref{fig:100Pack_bar}, we show the results for a pedestrian user with the same QoS requirements as previously described. In this scenario, we can observe that the random forest approaches the optimal solution both in terms of completed tasks and costs. Also Min Multi-x has similar performance. This is because, for a pedestrian user moving at $3$\,Km/h, the channel condition at $k=1,...,K$, does not vary a lot with respect to the measurements available at $k=0$. Note that, the results for $\beta=0$ and $\beta=0.5$ are the same. Namely, in this scenario, the random forest is able to obtain a better estimation of the optimal solution $\hat{z}$ and related probabilities $\hat{p}^j$. Finally, in Fig.~\ref{fig:100_Train_Trees}, we show the fraction of completed tasks for a vehicle user, while varying the number of trees that compose the random forest, and the training samples on which the model is trained. We can observe that the fraction of completed tasks is an increasing function of the number of training samples. Moreover, it is also possible to improve the performance by increasing the number of trees. This is more clear for larger training sets, whereas for fewer training samples increasing the number of trees does not provide significant gains. Namely, a large random forest provides a better estimation of the probabilities $\hat{p}^j$ and suffers less of overfitting~\cite{ISL}, but it increases the complexity and the duration of the training phase.

%

\section{Conclusion}
\label{sec:Conc}
In this work, we have proposed a learning-based solution for link scheduling in low-band assisted mm-wave multi-connectivity scenarios. This model is able to learn the optimal link scheduling solution for several user types and QoS requirements in order to maximize the number of completed tasks, minimize failed transmissions, and offload low frequency bands. Our proposed solution is a modified random forest classifier for which the predicted scheduling strategy is selected only if the estimated probability of being the optimal solution is higher than a threshold $\beta$, otherwise a fallback solution is used. The value of $\beta$ affects the tradeoff between allocated resources, costs and completed tasks.

We have shown that the proposed solution is able to outperform the baseline methods. Namely, the random forest is able to increase significantly  the amount of completed tasks with respect to the Min Multi-x solution with lower failed transmissions. Moreover, by increasing the value of $\beta$, the fraction completed tasks is further improved, with a slightly increase in the low-band usage and failed transmissions. Furthermore, the modified random forest is able to approach the performance of genie-aided and Greedy Multi-x solutions with a lower cost than the latter. Future work will investigate the impact on the performance of $T_{m}$ and the possibility to extend the proposed solution to the multi-user case.


\section*{Acknowledgment}
This work has received funding from the European Union's Horizon 2020 research and innovation programme under the Marie Sklodowska-Curie grant agreement No. 643002.





%


\bibliographystyle{IEEEtran}
\bibliography{ref}

\begin{thebibliography}{10}
\providecommand{\url}[1]{#1}
\csname url@samestyle\endcsname
\providecommand{\newblock}{\relax}
\providecommand{\bibinfo}[2]{#2}
\providecommand{\BIBentrySTDinterwordspacing}{\spaceskip=0pt\relax}
\providecommand{\BIBentryALTinterwordstretchfactor}{4}
\providecommand{\BIBentryALTinterwordspacing}{\spaceskip=\fontdimen2\font plus
\BIBentryALTinterwordstretchfactor\fontdimen3\font minus
  \fontdimen4\font\relax}
\providecommand{\BIBforeignlanguage}[2]{{%
\expandafter\ifx\csname l@#1\endcsname\relax
\typeout{** WARNING: IEEEtran.bst: No hyphenation pattern has been}%
\typeout{** loaded for the language `#1'. Using the pattern for}%
\typeout{** the default language instead.}%
\else
\language=\csname l@#1\endcsname
\fi
#2}}
\providecommand{\BIBdecl}{\relax}
\BIBdecl

\bibitem{BLO}
\relax M. {Xiao}~et al., ``Millimeter wave communications for future mobile
  networks,'' \emph{IEEE Journal on Selected Areas in Communications}, vol.~35,
  no.~9, pp. 1909--1935, Sept. 2017.

\bibitem{OurRel}
C.~{Tatino}, N.~{Pappas}, I.~{Malanchini}, L.~{Ewe}, and D.~{Yuan}, ``On the
  benefits of network-level cooperation in millimeter-wave communications,''
  \emph{IEEE Transactions on Wireless Communications}, vol.~18, no.~9, pp.
  4408--4424, Sept. 2019.

\bibitem{Meas}
\relax T. S. {Rappaport}~et al., ``Overview of millimeter wave communications
  for fifth-generation ({5G}) wireless networks - with a focus on propagation
  models,'' \emph{IEEE Transactions on Antennas and Propagation}, vol.~65,
  no.~12, pp. 6213--6230, Dec. 2017.

\bibitem{Danish}
D.~Aziz, J.~Gebert, A.~Ambrosy, H.~Bakker, and H.~Halbauer, ``Architecture
  approaches for {5G} millimetre wave access assisted by {5G} low-band using
  multi-connectivity,'' in \emph{IEEE Globecom Workshops}, Dec. 2016.

\bibitem{RLH}
Y.~{Sun}, G.~{Feng}, S.~{Qin}, Y.~{Liang}, and T.~P. {Yum}, ``Reinforcement
  learning based handoff for millimeter wave heterogeneous cellular networks,''
  in \emph{IEEE Global Communications Conference}, Dec. 2017.

\bibitem{NNH}
A.~{Alkhateeb}, I.~{Beltagy}, and S.~{Alex}, ``Machine learning for reliable
  mmwave systems: Blockage prediction and proactive handoff,'' in \emph{IEEE
  Global Conference on Signal and Information Processing}, Nov. 2018.

\bibitem{MLB}
A.~{Klautau}, P.~{Batista}, N.~{González-Prelcic}, Y.~{Wang}, and R.~W.
  {Heath}, ``{5G} {MIMO} data for machine learning: Application to
  beam-selection using deep learning,'' in \emph{Information Theory and
  Applications Workshop (ITA)}, Feb. 2018, pp. 1--9.

\bibitem{DLCB}
A.~{Alkhateeb}, S.~{Alex}, P.~{Varkey}, Y.~{Li}, Q.~{Qu}, and D.~{Tujkovic},
  ``Deep learning coordinated beamforming for highly-mobile millimeter wave
  systems,'' \emph{IEEE Access}, vol.~6, pp. 37\,328--37\,348, Jun. 2018.

\bibitem{Our1}
C.~{Tatino}, I.~{Malanchini}, N.~{Pappas}, and D.~{Yuan}, ``Maximum throughput
  scheduling for multi-connectivity in millimeter-wave networks,'' in
  \emph{16th International Symposium on Modeling and Optimization in Mobile, Ad
  Hoc, and Wireless Networks (WiOpt)}, May 2018.

\bibitem{ISL}
T.~Hastie, R.~Tibshirani, and J.~Friedman, \emph{The Elements of Statistical
  Learning -- Data Mining, Inference, and Prediction}.\hskip 1em plus 0.5em
  minus 0.4em\relax Springer--Verlag New York, 2009.

\bibitem{UPM}
M.~{Giordani}, M.~{Polese}, A.~{Roy}, D.~{Castor}, and M.~{Zorzi}, ``A tutorial
  on beam management for {3GPP} {NR} at mmwave frequencies,'' \emph{IEEE
  Communications Surveys Tutorials}, vol.~21, no.~1, pp. 173--196, First
  quarter 2019.

\bibitem{MCRT}
A.~{Wolf}, P.~{Schulz}, D.~{\"Ohmann}, M.~{Dörpinghaus}, and G.~{Fettweis},
  ``Rate-reliability tradeoff for multi-connectivity,'' in \emph{IEEE Wireless
  Communications and Networking Conference}, Apr. 2018.

\bibitem{CHM}
\relax M. R. {Akdeniz}~et al., ``Millimeter wave channel modeling and cellular
  capacity evaluation,'' \emph{IEEE Journal on Selected Areas in
  Communications}, vol.~32, no.~6, pp. 1164--1179, Jun. 2014.

\bibitem{MET}
J.~F. Monserrat and M.~F. (Editors), ``Mobile and wireless communications
  enablers for the twenty-twenty information society ({METIS-I}), deliverable
  d6.1, simulation guidelines,'' Tech. Rep., 2013.

\bibitem{ANT}
A.~{Awada}, A.~{Lobinger}, A.~{Enqvist}, A.~{Talukdar}, and I.~{Viering}, ``A
  simplified deterministic channel model for user mobility investigations in
  {5G} networks,'' in \emph{IEEE International Conference on Communications},
  May 2017.

\bibitem{scikit}
\relax F. Pedregosa~et al., ``Scikit-learn: Machine learning in {P}ython,''
  \emph{Journal of Machine Learning Research}, vol.~12, pp. 2825--2830, 2011.

\end{thebibliography}

\end{document}